\documentclass{elsart}
\usepackage{natbib}
\usepackage{graphicx}
\def\ETi{E_{{\rm T}_i}}
\def\gg{\ifmmode{\gamma\gamma}\else{$\gamma\gamma$}\fi}
\def\kp{\ifmmode{k_{\perp}}\else{$k_{\perp}$}\fi}
\def\as{\ifmmode{\alpha_{\rm s}}\else{$\alpha_{\rm s}$}\fi}
\def\sqee{\ifmmode{\sqrt{s}_{\rm ee}}\else{$\sqrt{s}_{\rm ee}$}\fi}
\def\xg{{\ifmmode{x_{\gamma}}\else{${x_{\gamma}}$}\fi}}
\def\epem{{\ifmmode{{\mathrm e}^{+}{\mathrm e}^{-}}\else{${\mathrm e}^{+}{\mathrm e}^{-}$}\fi}}
\def\egam{{\ifmmode{{\mathrm e}{\gamma}}\else{${\mathrm e}{\gamma}$}\fi}}
\def\gamp{{\ifmmode{{\gamma}{\mathrm p}}\else{${\gamma}{\mathrm p}$}\fi}}
\begin{document}
%
%
\runauthor{Thorsten Wengler, Albert De Roeck}
\begin{frontmatter}
\title{The gluon content of the photon from di-jet production at the 
\boldmath {\gg}-collider \unboldmath}
\author[CERN]{Thorsten Wengler}
\author[CERN]{Albert De Roeck}
\address[CERN]{CERN, EP divison, 1211 Geneva 23, Switzerland}
%
%
\begin{abstract}
A study of di-jet measurements at a future photon-photon collider
is reported. The sensitivity to extract the gluon distribution
is discussed. The results are 
 compared  with  calculations for a 
linear {\epem} collider.

\end{abstract}
%
%
\begin{keyword}
di-jets; Photon-Photon collider, gluon density in the photon
\end{keyword}
\end{frontmatter}
%
%
\section{Introduction}
A future linear {\epem} collider in the few hundred GeV centre of mass
(CMS)  energy  range,  offers   the  opportunity  for  a  high  energy
{\gg}-collider~\cite{bib-telnov},  i.e.   collisions  of  high  energy
photon  beams  with  a  high  luminosity.  This  can  be  achieved  by
Compton-backscattering of photons off a laser beam on the electron and
positron  beams  of  the   {\epem}  collider.  Under  the  appropriate
kinematical  and  geometrical conditions  a  narrow  spectrum of  high
energy photons is produced, which can reach up to 80\% of the original
{\epem} beam energy.

\begin{figure}[ht]
\begin{center}
\includegraphics[width=0.45\textwidth]{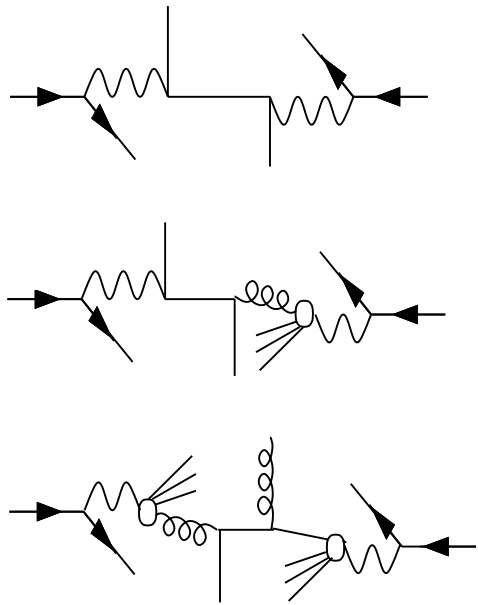}
\end{center}
\caption{ Examples of 
Leading Order production diagrams (from top to bottom): direct, single
resolved and double resolved processes.}
\label{fig:feyn}
\end{figure}

The  QCD  process of  two-photon  jet  production  has a  large  cross
section, thus the measurements will be dominated by systematics.  Jets
can be  produced in  hard partonic scattering  in direct  and resolved
processes.  For  resolved processes the  photon converts first  into a
hadron  with   vector  meson  quantum   numbers,  or  splits   into  a
quark-antiquark pair. The collision  of two resolved photons is called
a double  resolved process. A photon  can also couple  directly to the
quarks of the resolved second photon (called single resolved process),
or to  another bare photon via  a quark loop  (called direct process).
Examples of these processes are shown in Fig.~\ref{fig:feyn}.

The resolved processes involve the hadronic component from one or both
photons  and are  thus  sensitive  to the  hadronic  structure of  the
photon.   Measurements of the  hadronic structure  of the  photon have
been made in {\egam} scattering  at {\epem} colliders over the last 20
years~\cite{bib-nisius}.    These   measurements   are   predominantly
sensitive to the quark content of the photon, although they also start
to constrain the gluon content of the photon via scaling violations at
low values of {\xg}.  Measurements  of jet cross sections on the other
hand are directly sensitive to both the quark and gluon content of the
photon  and  can  thus,  in  combination  with  results  from  {\egam}
scattering,  be  used  to   measure  the  gluon  distribution  in  the
photon.  Recently studies of  di-jet events  in {\gg}  interactions at
LEP~\cite{bib-djopold,bib-twichep}    and    {\gamp}   photoproduction
interactions at  HERA~\cite{bib-hera} have shown  a direct sensitivity
to the gluon distribution in the photon.

In  this paper  a study  is presented  of di-jet  cross sections  at a
photon  collider (PC),  and the  sensitivity to  the structure  of the
photon is investigated. The results are compared with expectations for
a linear {\epem} collider.

\section{Monte Carlo Study}

The  study is  made  with  the Monte  Carlo  Generator program  PHOJET
(version 1.12)~\cite{bib-phojet} which contains all leading order (LO)
diagrams  complemented  by  parton  showers to  emulate  higher  order
contributions.  Hadronization of the parton final state is provided by
the  JETSET~\cite{bib-pythia} program.  The  transverse energy  of the
jets  provides  a  hard   scale  which  regularizes  perturbative  QCD
calculations.    Jets   are    reconstructed   using   the   inclusive
{$k_{\perp}$}       clustering       algorithm       as       proposed
in~{\cite{bib-ktclus}}. In this algorithm the distance measure between
any pair of  objects $\left\{i,j\right\}$ to be clustered  is taken to
be   $d_{ij}   =   \mathrm{min}\left(E^{2}_{T,i},E^{2}_{T,j}   \right)
\left(R^{2}_{i,j}/R^{2}_{0}\right)$      with      $R^{2}_{i,j}      =
\left(\Delta\eta_{ij}\right)^{2}  + \left(\Delta\phi_{ij}\right)^{2}$.
Throughout  this   analysis  we  set  $R^{2}_{0}   =  1$.   $E_{T,i}$,
$\eta_{i}$, and $\phi_{i}$  are the transverse energy, pseudorapidity,
and  azimuthal   angle  of  the   $i$-th  object  in   the  laboratory
center-of-mass system.  The transverse energy $\ETi$ of  an object $i$
is defined  relative to the $z$  axis of the  detector. The clustering
starts from the smallest  value $d_{\mathrm{min}}$ of the combined set
of all  $d_{ij}$ and all $d_{i} =  E^{2}_{T,i}$. If $d_{\mathrm{min}}$
belongs to the subset $d_{ij}$, the two objects $i$ and $j$ are merged
into  a new  object using  the $E_{T}$  recombination scheme.  In this
scheme  the properties  of the  merged particle  are computed  as $E^{
\prime}_{T}  = E_{T,i} +  E_{T,j}$, $\eta^{  \prime} =  \left( E_{T,i}
\eta_{i} +  E_{T,j} \eta_{j} \right)  / E^{ \prime}_{T}$,  and $\phi^{
\prime} =  \left( E_{T,i}  \phi_{i} + E_{T,j}  \phi_{j} \right)  / E^{
\prime}_{T}$. If $d_{\mathrm{min}}$ belongs to the subset $d_{i}$, the
object is  added to the list  of jets and removed  from the clustering
list.  The procedure  is  finished when  no  objects are  left in  the
clustering list.

Realistic  detector   cuts,  imposed  by  the   TESLA  detector  study
group~\cite{bib-tesladet}  are  applied  in  the  Monte  Carlo  study.
Unless otherwise specified jets  are required to have a pseudorapidity
$\eta$ in the range $|\eta| < 2.5$.

The kinematics of the two jets is used to estimate the fraction of the
photon momentum  participating in the  interaction, {\xg}, which  is a
sensitive probe of  the structure of the photon.   A variable, defined
at the hadronic final state, is
$$ x^{\pm}_{\gamma}=\frac{\Sigma_{jets}(E\pm p_z)}
{\Sigma_{hadrons}(E\pm p_z)}$$ 
and is  closely related to the true  partonic {\xg} variable.  Here
$E$   and  $p_z$   denote   the  energy   and  longitudinal   momentum
respectively.  Direct  processes are  characterized by having  a {\xg}
value close to one.

\begin{figure}[ht]
\begin{center}
\includegraphics[width=0.95\textwidth]{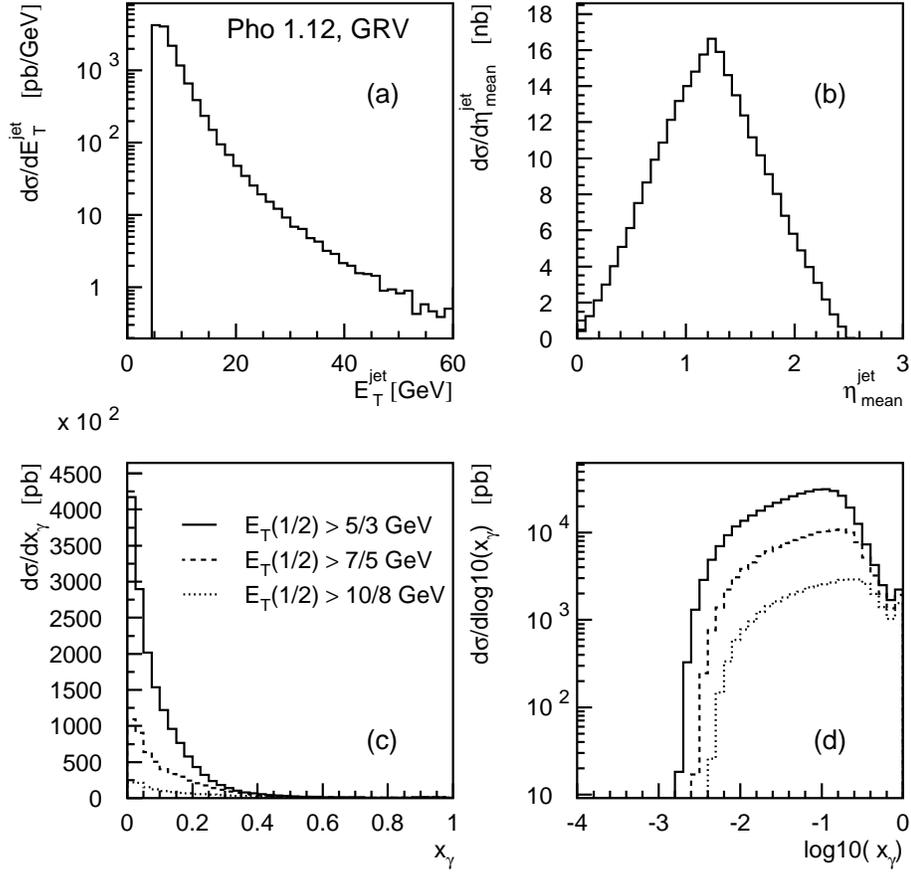}
\end{center}
\caption{ The di-jet cross section as a function of 
$E_T$ and $\eta^{jet}_{mean} = (\eta^{jet1}+\eta^{jet2})/2$ and
{\xg}. The latter is shown on a linear and logarithmic scale, and
 for increasing thresholds of the jet transverse energy.}
\label{fig:etcut}
\end{figure}

A linear \epem collider with a  CMS energy of 500 GeV is assumed.  The
expected luminosity of TESLA is 200 fb$^{-1}$/year, and for the photon
collider about 10 times lower (for  the high energy part of the photon
spectrum).

\section{Results}

Fig.~\ref{fig:etcut}  shows  the di-jet  cross  section  for a  photon
collider, as a function of $E_T$ of the jets and the average $\eta$ of
the jet pair.  The {\xg} distribution is shown  as function of minimum
$E_T$ thresholds for the jets (each event gives two entries: ${\xg}^+$
and ${\xg}^-$). E.g. the entry 5/3 GeV means that the first jet has an
$E_T$  above  5  GeV, while  the  second  jet  has  an $E_T$  above  3
GeV.  Clearly lower  thresholds allow  to reach  lowest  {\xg} values,
which are  of particular  interest. In this  region the  gluon induced
processes are anticipated  to dominate, and moreover a  strong rise of
the gluon  density is predicted at  small $x$ from  models mirrored to
the evolution dynamics of proton parton densities. For the photon such
a rise 
at small  $x$ has not  been unambiguously observed yet.   The proposed
measurements  reach  an  order  of  magnitude further  down  in  {\xg}
compared   to   current   measurements    from   jets   at   LEP   and
HERA~\cite{bib-djopold,bib-twichep,bib-hera}.

It is important  to note that also for the  highest $E_T$ cuts studied
there is still a good sensitivity  to parton densities at low $x$. The
final choice of the $E_T$ thresholds will depend on the acceptance and
background conditions and on the control of theoretical uncertainties,
such as  underlying event effects,  which may spoil the  extraction of
the gluon distribution from the  data. Generally the cuts are expected
to be in  the range given in this figure.  The  cross sections for all
threshold  combinations  are large,  hence  the  measurement of  these
processes will  be systematics dominated.   Unless stated differently,
the  results below  are for  the (5/3)  GeV threshold  pair,  which is
presently used in jet studies at LEP.
\begin{figure}[ht]
\begin{center}
\includegraphics[width=0.95\textwidth]{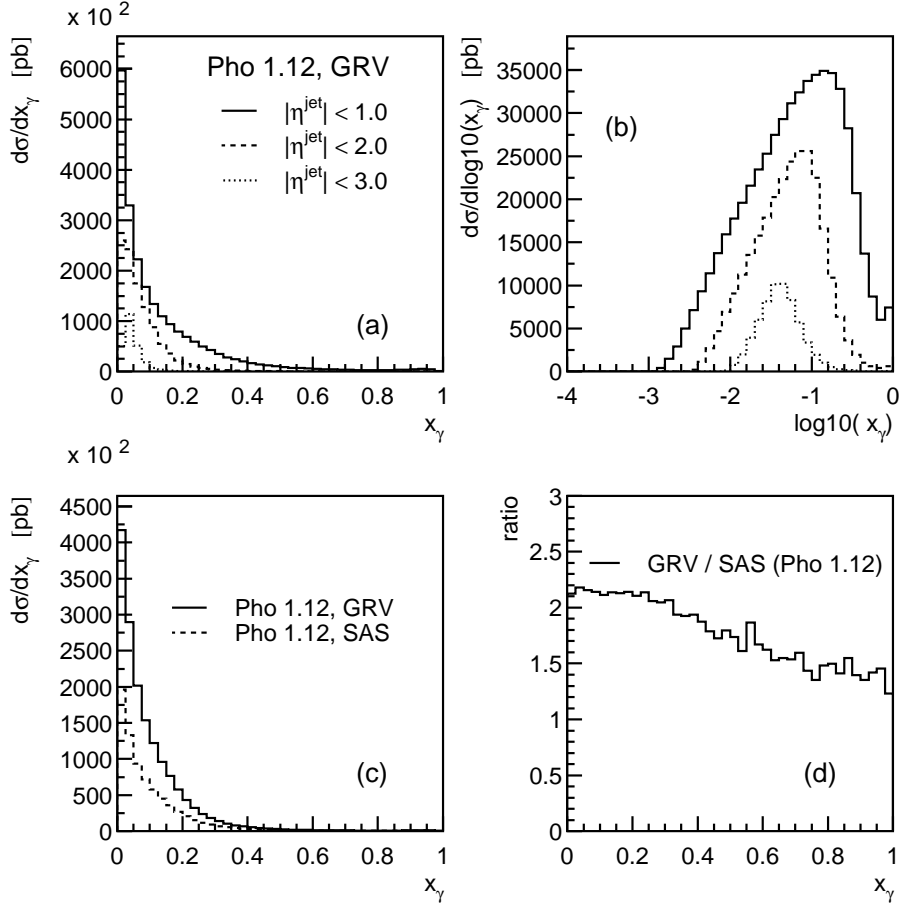}
\end{center}
\caption{ The effect of the $\eta$-acceptance on the reach in {\xg}
  (a,b). (c) shows the di-jet cross section as a function of {\xg} 
  for two different parton distribution functions, GRV and SaS. In 
  (d) the ratio of the  cross sections is shown.} 
\label{fig:eta}
\end{figure}
\begin{figure}[ht]
\begin{center}
\includegraphics[width=0.95\textwidth]{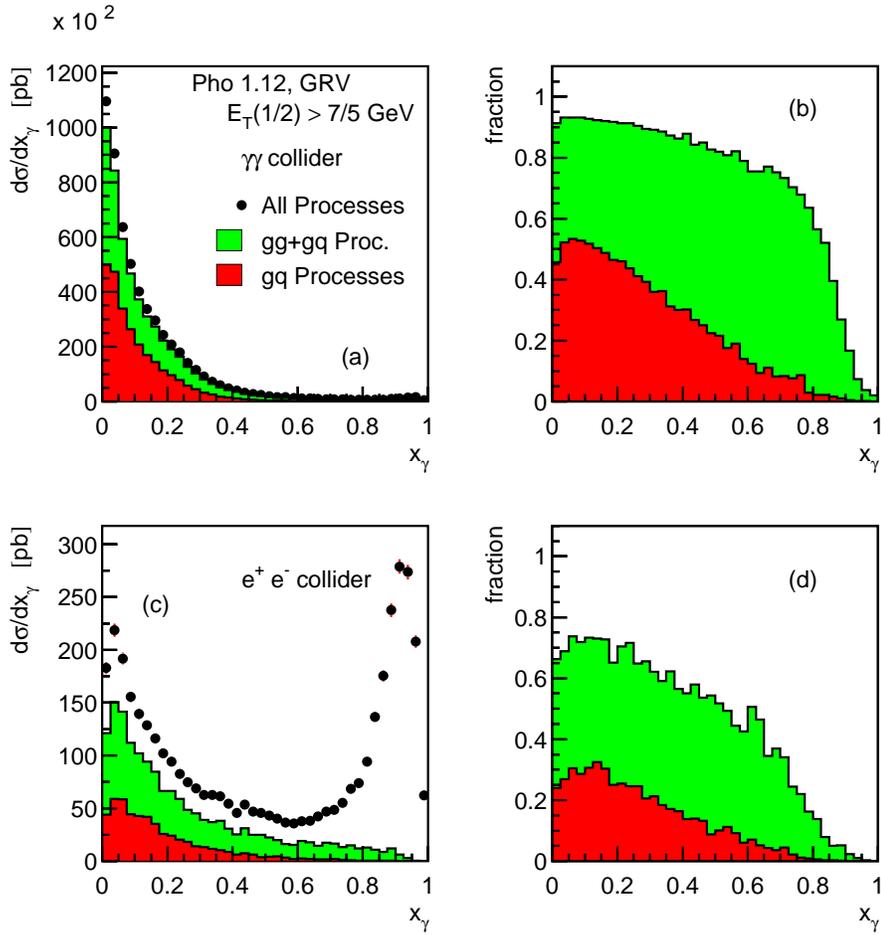}
\end{center}
\caption{ The contribution of gluon initiated processes to the di-jet
  cross section as a function of {\xg}. The case of the {\gg}-collider
  in the upper plots is compared to the {\epem}-option in the
  lower plots. On the
  left hand side the relative contributions are shown.}
\label{fig:gluon}
\end{figure}
Fig.\ref{fig:eta}(a,b)  shows the  effect from  the  restricted $\eta$
range for jet  measurements in a future detector  on the low-$x$ reach
of {\xg}.  The effect is  large, and better $\eta$ coverage will allow
to reach smaller values in {\xg}.

Fig.\ref{fig:eta}(c,d) show the sensitivity  of the cross section as a
function  of  {\xg} to  two  different  parton  distributions for  the
photon, \mbox{GRV LO}~\cite{bib-grv} and \mbox{SAS 1D}~\cite{bib-sas},
which are both 
consistent with  present day measurements of the  photon structure. At
low $x$ the differences in cross section are as large as a factor two,
well within the sensitivity expected of the measurement.

In   Fig.\ref{fig:gluon}(a,b)  the   contribution  of   gluon  induced
processes, compared  to all  processes, is shown  for (7/5)  GeV $E_T$
thresholds.  A  small smeared peak  from direct contributions  is seen
close to  {\xg} =1. However  in most of  the region the  processes are
dominated by parton scattering involving gluons, and thus sensitive to
the  gluon  distribution in  the  photon.  At  small {\xg}  the  gluon
contribution amounts to 90\%.

Fig.~\ref{fig:gluon}(c,d)  shows  the same  distributions  but for  an
{\epem} collider,  where two  photon interactions result  from photons
emitted  from the  beam leptons  described by  a Weiz\"acker-Williams
energy distribution.  Hence the  energy of the  photons is  in general
much smaller than that of the photons from a photon collider, based on
the same {\epem}  machine.  The cross sections at  an {\epem} collider
are much  smaller, thus  loosing the benefit  of a  higher luminosity.
Furthermore,  the   quark  induced  processes   have  relative  larger
contributions at  an {\epem} collider, blurring the  extraction of the
gluon  distribution from  the jet  measurement much  more than  in the
photon collider case.

\section{Summary}

Di-jet  production  at a  future  photon  collider  has been  studied,
including restrictions  imposed by a possible  detector, and discussed
in terms  of the  {\xg} variable  which is a  measure of  the hadronic
structure  of  the  photon.  With presently  anticipated  experimental
restrictions, values  of {\xg}  down to a  few times $10^{-3}$  can be
reached. At low  {\xg} the cross section is  predicted to be dominated
by the  gluon content of the photon,  and can thus be  used to extract
the gluon  distribution.  The photon collider  has several advantages,
such as  large cross  sections and better  dominance of  gluon induced
processes, over a {\epem} collider.

%
%
\bibliography{djets}
%

\end{document}